\documentclass{appolb}
\usepackage{graphicx}

\begin{document}
\title{The Roper Resonance: Still Controversial 60 Years Later%
}
\author{Anthony W.~Thomas
\address{CSSM, Department of Physics, The University of Adelaide, SA 5005 Australia}
}
\maketitle
\begin{abstract}
Few baryon resonances have generated as much discussion, even controversy, as the first positive parity excited state with nucleon quantum numbers. We re-examine the issue using insight gained from lattice QCD, complemented by Hamiltonian effective field theory. In doing so, we also examine the distinction between a state that can be naturally described as a quark model state and one that is dynamically generated.
\end{abstract}
  
\section{Introduction}
It is a great pleasure to dedicate this article to Dave Roper on his ninetieth birthday. Discovered more than 60 years ago~\cite{Roper:1964zza}, the Roper resonance with nucleon quantum numbers but roughly 500 MeV higher in energy, has provided challenges ever since. Within a simple harmonic oscillator model (SHM) for the confining potential, which has been very widely used~\cite{Isgur:1978wd,Isgur:1977ef}, one expects the first excited state to have negative parity with the first positive parity state at roughly double that excitation energy. On the other hand, experiment reveals that the Roper is in fact some 85 MeV {\em below} the first negative parity excited state.

Later calculations involving a Coulomb-like hyper-central confining potential~\cite{Giannini:2015zia} suggested a resolution of this problem. An alternate approach has also been used based upon a relativistic quark model with a Y-shaped confining potential motivated by lattice simulations~\cite{Capstick:1986ter}. On the other hand, the MIT bag model~\cite{Chodos:1974pn}, which is fully relativistic and at least crudely incorporates the non-perturbative structure of the QCD vacuum, confines the quarks in a spherical cavity and suggests a very similar ordering to that of the SHM. The Nambu-Jona-Lasinio (NJL) model has also been applied to baryon excited states, with the conclusion that the Roper is indeed a positive parity excitation of a three-quark system~\cite{Segovia:2015hra}.

It is apparent that the diversity of models for the mass of this state does not allow us to draw firm conclusions about its nature. Even more uncertainty is associated with proposals that the Roper is not predominantly a three-quark state but is dynamically generated through strong meson-baryon scattering~\cite{Krehl:1999km}. 

On the experimental side there have been important advances~\cite{Burkert:2025coj}, with the resonance studied using both photo-~\cite{Dutz:1996uc} and electro- production~\cite{Thiel:2022xtb,CLAS:2008roe,CLAS:2009ces,Aznauryan:2011qj} reactions, as well as the traditional pion-nucleon scattering~\cite{Kelly:1979uf,Koch:1980ay}.

In this brief review we take a new look at the Roper resonance in the light of recent lattice QCD calculations, complemented by Hamiltonian effective field theory. 

\section{Hints from Lattice QCD}
We are familiar with Sherlock Holmes conclusions based upon the fact that ''the dog didn't bark''. This is more or less what started a new interpretation of the Roper resonance. For many years lattice QCD calculations of this excited state failed to produce a state anywhere near 1.45 GeV. Rather, they yielded masses far above the observed state; typically 1.8 to 1.9 GeV, rather than 1.45~\cite{Kiratidis:2016hda,Mahbub:2010rm,Edwards:2011jj}. These calculations used three-quark interpolating fields (sources and sinks) and therefore should have strongly excited a genuine quark model state, but they did not. While the mass found was inconsistent with the Roper resonance, study of the wave function of the state on the lattice did reveal the expected node of a genuine 2s excited state~\cite{Roberts:2013ipa}.

Of course, in principle, all possible configurations with the quantum numbers of the state under investigation should be excited, regardless of the interpolating field used. However, in practice, a local operator has a rather small overlap with a continuum state (eg., a meson-baryon system) that is spread over the entire lattice. It was only when the first calculation by the Graz group~\cite{Lang:2016hnn}, which included explicit pion-nucleon and $\sigma$-nucleon interpolating fields, that the lattice calculations produced a resonance at the observed mass. These authors immediately recognized that their results were consistent with the earlier study by the CSSM group~\cite{Kiratidis:2016hda,Leinweber:2015kyz}, which suggested that the Roper was dynamically generated by meson-baryon scattering, rather than being predominantly a three-quark (or quark model) state.

In concluding this section we emphasize that lattice QCD is not a model. It allows us to calculate the non-perturbative structures that are generated by QCD itself. That is why the results of Lang {\em et al.} in Graz were so important.

\section{Dynamically generated states}
As there is considerable confusion in the literature over the distinction between quark states and dynamically generated states, a few words of explanation may be helpful.

In the real world there is no such thing as a pure quark state with a mass above the meson baryon thresholds with the same quantum numbers. All states of this kind will decay and so some of the strength at the resonance pole will be associated with those continuum states; they are strictly not eigenstates of the QCD Hamiltonian. Of course, on the lattice we do calculate the eigenstates of the QCD Hamiltonian in a finite volume and all such states are stable. If we include meson-baryon interpolating fields in the lattice calculation, these eigenstates will be superpositions of three quark and meson baryon states. This makes it very clear where the confusion may arise.

In order to appreciate how one might interpret such problems let us briefly recall the first discussion of this issue. This involves the $\Delta(1232)$ resonance, which lies above the pion nucleon threshold and has a width of roughly 120 MeV. Before quarks were invented this state was understood in the Chew-Low model~\cite{Chew:1955zz} as being dynamically generated by multiple scattering through the so-called crossed-Born diagram, which is strongly attractive. In order for this to work, it is essential that the form-factor, or high momentum cut-off, at the pion nucleon vertex be hard. That is, it must not suppress the emission of high momentum pions. After the invention of the MIT bag model~\cite{Chodos:1974je}, this was initially interpreted by Gerry Brown and collaborators~\cite{Brown:1979ui} as support for their argument that the bag radius must be very small.

The fundamental role of chiral symmetry in QCD~\cite{Thomas:2025wlj} required important changes to the MIT bag model. In particular, it demanded that pions couple to the confined quarks at~\cite{Miller:1979kg}  or near~\cite{Thomas:1981ps,Thomas:1982kv} the bag boundary. As with any emission of absorption process from an extended object, there must therefore be a form factor which suppresses the emission of high energy pions in inverse proportion to the size of the source. While the existence of such form factors appears to be anathema to many working in effective field theory, they are the natural result of simple physics. This form factor controls the strength of the Chew-Low mechanism. 

However, in the MIT bag model there is a three quark $\Delta$ eigenstate, separated in mass primarily~\cite{Young:2002cj} by one-gluon-exchange and with essentially the same bag radius as the nucleon. As a result, the same form factor controls the coupling of the pion to the three quark $\Delta$ found in the MIT bag model and to the nucleon. One is {\em not} free to vary those independently. It is then natural to ask, why do we not find two  $\Delta$ resonances? 

Théberge, Miller and Thomas used the Hamiltonian derived in the cloudy bag model, in which the pion was coupled to the confined quarks in a way to ensure chiral symmetry, to study pion nucleon scattering in the $\Delta$ region~\cite{Theberge:1980ye}. The unambiguous conclusion was that the observed resonance was dominated by the decay of the $\Delta$ bag state. This carried roughly 80\% of the strength at the resonance pole. Because the underlying theory did not allow arbitrary variations in the form factors, the radius of the bag found in the fit, $0.72 \pm 0.14$ fm, naturally suppressed the contribution through the Chew-Low mechanism. Indeed, with the form factor fixed and the coupling to the $\Delta$ turned off, the theory yields no resonance at all.
\begin{figure}[htb]
\centerline{%
\includegraphics[width=12.5cm]{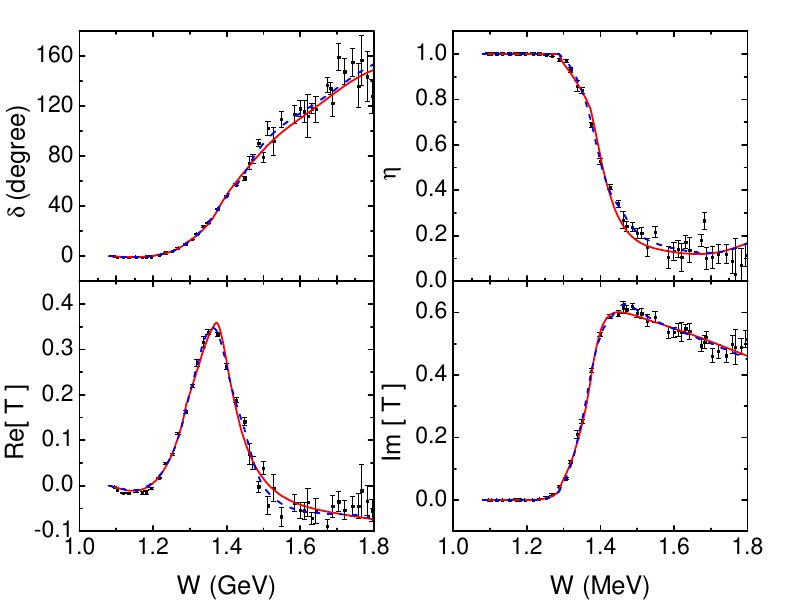}}
\caption{Pion nucleon phase shifts and inelasticities calculated in the $P_{11}$ channel for the two theoretical models described in the text. The quality of the fit is indistinguishable for these two cases -- from Ref.~\cite{Wu:2017qve}.}
\label{Fig:1}
\end{figure}

This theoretical treatment was effectively the first application of what we now call Hamiltonian effective field theory (HEFT) to resonance structure in QCD; albeit without the additional constraint of data from lattice QCD. It provides a text book example~\cite{Thomas:2001kw} of how one might identify a quark model state. One must have an underlying Hamiltonian which is consistent with chiral symmetry and the postulated quark structure. One must fit the available scattering data and, when it is available, the lattice data on the corresponding finite volume. If the dominant strength at the resonance pole is associated with the quark structure it is a quark model state. However, if the strength associated with the quark state is small, the resonance must be regarded as dynamically generated.

\section{HEFT for the Roper resonance}
By late 2015 we not only had the extensive lattice QCD studies by the CSSM~\cite{Roberts:2013oea}, JLab~\cite{Edwards:2011jj} and Cyprus~\cite{Alexandrou:2014mka} groups, but also the work of Lang {\em et al.}~\cite{Lang:2016hnn}, which we have already mentioned. This enabled Wu {\em et al.}~\cite{Wu:2017qve} to conduct a careful analysis of this lattice data, as well as the experimental phase shifts, within HEFT. These authors considered two scenarios; one in which the quark state had a high mass (of order 2 GeV) and the other where it was much nearer (around 1.7 GeV) the observed resonance. As we see in Fig.~\ref{Fig:1}, the quality of the fit to the phase shifts is equally good in these two cases. 
\begin{figure}[htb]
\centerline{%
\includegraphics[width=12.5cm]{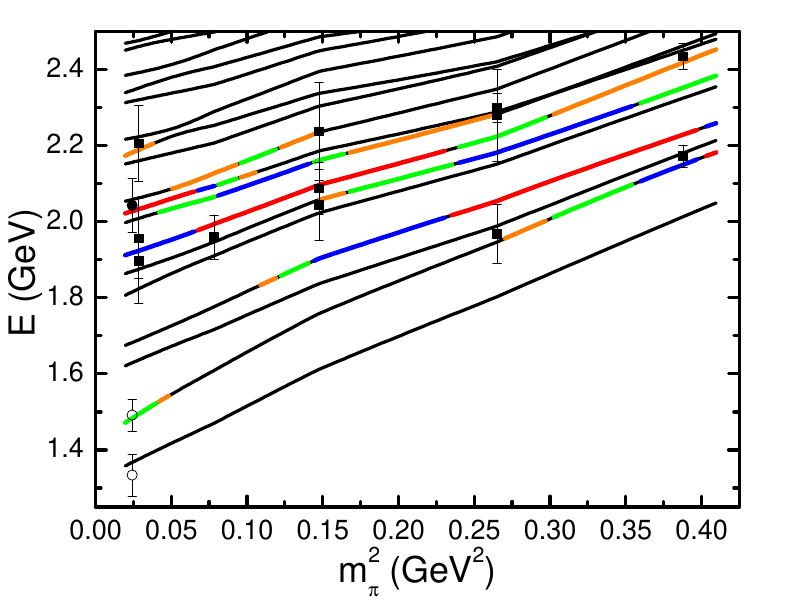}}
\caption{Comparison of the finite volume energy levels calculated in a scenario where the bare, or quark model, state is around 2 GeV, with the eigenenergies reported by the CSSM group (solid boxes) and Lang {\em et al.} (open circles). The color coding indicates those states with a large quark model component, which are therefore most likely to be seen in lattice simulations with a three-quark interpolating field. It is clear that the states seen on the lattice match the expectations very well -- from Ref.~\cite{Wu:2017qve}.}
\label{Fig:2}
\end{figure}

In Fig.~\ref{Fig:2} we show a comparison of the eigenstates calculated using HEFT in a finite volume, compared with those found on a lattice of the same size, for the case where the mass of the bare state was large. The color coding illustrates those states in which there is a large bare state component and therefore should be expected to be seen in a lattice calculation using a three-quark interpolating field -- as in the CSSM simulations indicated by the solid squares. The open circles corresponding to the results of Lang {\em et al.}, which we recall included $\sigma$-nucleon and pion nucleon interpolating fields, correspond to levels with only a small three-quark component. It is the highest of those two levels which corresponds to the observed resonance energy. 

On the other hand, in the scenario where the bare state has a lower mass, there is no correspondence between the states expected to be seen with a three quark interpolating field and those actually found in the corresponding lattice QCD calculations.

These results suggest unambiguously that the observed Roper resonance at 1.45 GeV is predominantly a dynamically generated state, with only a very small three quark component. The actual 2s excitation of the three quark system then has a mass around 1.9-2 GeV. As noted earlier, this assignment is consistent with the quark wave function measured by the CSSM lattice group.

\section{Concluding remarks}
The analysis outlined in the previous section strongly suggests that the Roper resonance observed at 1.45 GeV is not a three quark state but rather is dynamically generated through strong rescattering between the pion nucleon and $\sigma$ nucleon channels. 

Of course, the analysis of this fascinating resonance is not yet complete. Considerable emphasis has been put by a number of authors on the beautiful pion electro-production data for the Roper resonance, which shows a change in sign in the $A_{1/2}$ helicity amplitude at a momentum transfer around 0.6 GeV$^2$~\cite{Rodriguez-Quintero:2019yec,Burkert:2017djo}. It has been suggested that this may be related to the node in the 2s wave function; supposing that the Roper is indeed a 2s excitation. As we have explained, this does not appear to be consistent with the fact that lattice studies show a node in the wave function of the nucleon excited state found near 1.9-2 GeV, rather than in the mass region around the Roper resonance. It is therefore important that this data be studied in detail. For now, only photo-production has been calculated using the HEFT described earlier~\cite{Zhuge:2024iuw}, which correctly explains the lattice data and the scattering phase shifts.

Clearly, even after 60 years this resonance continues to provide us with profound new challenges.

\section*{Acknowledgments}
It is a pleasure to acknowledge particularly my extensive discussions and collaboration with Derek Leinweber on the issues discussed here. The collaboration with Zhan-Wei Liu, Waseem Kamleh and Jiajun Wu is also much appreciated. This work was supported by the University of Adelaide and by the Australian Research Council through Discovery Project DP230101791.


%
\end{document}